\documentclass[letter,twocolumn]{jpsj2}

\def\runauthor{Katsunori \textsc{Kubo} and Peter \textsc{Thalmeier}}

\title{Multiorbital Effects
  on Antiferromagnetism in Fe Pnictides}

\author{Katsunori \textsc{Kubo}$^{1,2}$ and Peter \textsc{Thalmeier}$^1$}
\inst{$^1$Max Planck Institute for Chemical Physics of Solids,
01187 Dresden, Germany\\
$^2$Advanced Science Research Center,
Japan Atomic Energy Agency,
Tokai, Ibaraki 319-1195}

\recdate{May 13, 2009;
  accepted June 16, 2009; published July 27, 2009}

\abst{
We apply a Hartree-Fock approximation to a two-orbital model
proposed for Fe pnictide superconductors.
It is found that the antiferromagnetic (AFM) order with the ordering vector $\mib{Q}=(\pi,0)$ is realized.
The AFM order appears simultaneously with ferro-orbital order,
the latter leads to a secondary lattice distortion.
We also investigate the influence of doping on the AFM order.
The size of the AFM moment changes continuously
for lightly doped cases,
but when the amount of doped carriers exceeds a certain value
the AFM state is suddenly destroyed.
We also show that Fermi surfaces remain and change significantly on doping
even in the AFM state.
This behaviour is explained
by considering the nesting due to the multi-sheet Fermi-surface structure
and multiorbital nature of the electronic bands
characteristic to Fe pnictides.
}

\kword{iron pnictides, magnetic order, orbital order, lattice distortion,
  Fermi surface, superconductivity}

\begin{document}
\maketitle

Since the discovery of superconductivity in LaFeAsO$_{1-x}$F$_x$
with a high transition temperature $T_c = 26$~K,~\cite{Kamihara}
extensive studies have been done on Fe pnictides.
The main interests on these materials
are not only on the high transition temperates
such as $T_c=55$~K in SmFeAsO$_{1-x}$F$_x$~\cite{Ren}
and $T_c=56$~K in Gd$_{1-x}$Th$_x$FeAsO,~\cite{Wang}
but also on the mechanism of the superconductivity.
The electronic structure is quasi-two-dimensional~\cite{Lebegue,Singh_La,Dong,Mazin,Xu,Singh_Ba}
and superconductivity occurs around the magnetic phase boundaries~\cite{Kamihara,Rotter_dope2,Kotegawa,Chu,Luetkens}
as in high-$T_c$ cuprates.
Such similarities suggest that
magnetism is probably playing an important role in the emergence of superconductivity,
and it is highly desirable to unveil the microscopic origin of
magnetism characteristic to Fe pnictides.

The magnetism in Fe pnictides is much different from that in cuprates.
In the latter, ordering vector of the antiferromagnetism is $(\pi,\pi)$,
while it is $(\pi,0)$ in Fe pnictides
in the unfolded Brillouin zone (BZ) with one Fe ion per unit cell.~\cite{Cruz,Zhao,Huang,Kaneko}
The undoped antiferromagnetic (AFM) states are metallic~\cite{Rotter_dope2,Kotegawa,Chu,Rotter_undope,Rotter_dope1,Sasmal}
in Fe pnictides while insulating in cuprates.
The AFM transition occurs at~\cite{Zhao,Huang,Kaneko,Tegel,Jesche}
or near~\cite{Cruz} the structural
transition temperature in Fe pnictides.

\begin{figure}
  \begin{center}
    \includegraphics[width=0.9\linewidth]{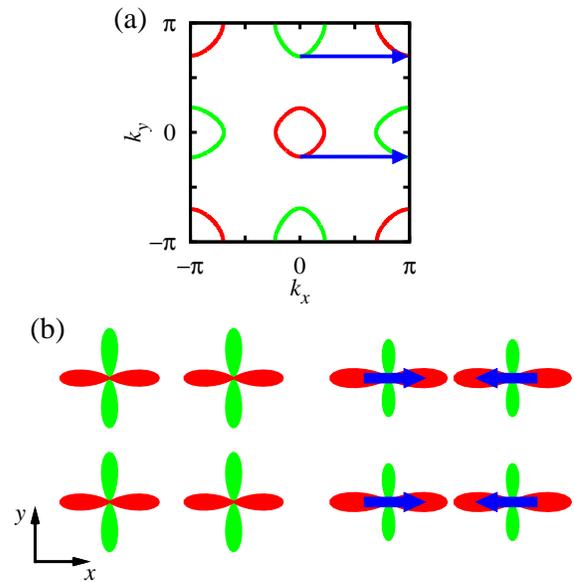}
  \end{center}
  \caption{\label{figure:schematic}
    (Color online)
    (a) Fermi surfaces of the two-orbital model~\cite{Raghu}
    in the unfolded BZ at electron number $n=2$ per site.
    The arrows indicate the nesting vector $\mib{Q}=(\pi,0)$.
    The hole surfaces locate around $(0,0)$ and $(\pi,\pi)$,
    and the electron surfaces locate around $(\pi,0)$ and $(0,\pi)$.
    (b) Schematic views of the orbital states in the normal state (left)
    and the AFM state accompanying ferro-orbital order (right).
    The orbital stretched along $x$ ($y$) axis represents
    the $d_{zx}$ ($d_{yz}$) orbital.
    The arrows represent the spin state, and
    the sizes of the orbitals indicate the occupancies of those orbitals.
  }
\end{figure}

Such differences in magnetism may originate
from the multiorbital electronic states and multi-sheet Fermi-surface (FS) structure
in Fe pnictides.~\cite{Lebegue,Singh_La,Dong,Mazin,Xu,Singh_Ba}
Indeed, the AFM order with $(\pi,0)$
due to nesting between hole and electron pockets
[see Fig.~\ref{figure:schematic}(a)]
has been suggested by using tight-binding models~\cite{Kuroki,Raghu,Daghofer}
and by band-structure calculations.~\cite{Dong,Mazin}
Yildirim~\cite{Yildirim} has shown that
the tetragonal lattice distortion occurs in the AFM state with $(\pi,0)$,
but not in the normal state.
In addition, the lattice distortion occurs neither
in an AFM state with $(\pi,\pi)$ nor in a ferromagnetic state.
Ran \textit{et al.}~\cite{Ran} have shown that
a full band gap does not open and the system remains metallic
even in the AFM state
from a topological view point of the multiorbital system.

In this Letter,
we show that such characteristic features on magnetism are explained even
in the simplest model, i.e., by a two-orbital model,
proposed for Fe pnictides~\cite{Raghu,Daghofer} by applying Hartree-Fock approximation.
While the Hartree-Fock approximation has already been applied
to the two-orbital model,\cite{Yu}
possibility of orbital order and effects of doping
have not been investigated yet.
They are main topics of the present study.
The two-orbital model cannot reproduce well the band structure
obtained with the density functional theory,
while a five-orbital model does well.~\cite{Kuroki}
In addition, $d_{xy}$ orbital, which is not included in the two-orbital model,
contributes some portions of the FSs.\cite{Zhang}
Thus, to describe some properties, all the five $d$-orbitals may be necessary.
However, the two-orbital model can reproduce
at least the characteristic FS topology in Fe pnictides
in the folded BZ,
and is enough for the purpose of the present Letter.
In this study, we take into consideration orbital order
on an equal footing with AFM order,
since they are closely related to each other in Fe pnictides.
Indeed, the AFM order with the ordering vector $\mib{Q}=(\pi,0)$
inevitably accompanies ferro-orbital (FO) order [schematically shown in Fig.~\ref{figure:schematic}(b)]
which results in a secondary orthorhombic distortion.
In addition, we investigate doping effects on the antiferromagnetism and
FSs reconstructed by the AFM order.
As shown in Fig.~\ref{figure:schematic}(a),
there are two kinds of nesting with the same nesting vector $\mib{Q}=(\pi,0)$,
i.e., between FSs around (0,0) and $(\pi,0)$,
and between FSs around $(0,\pi)$ and $(\pi,\pi)$.
The existence of the two kinds of nesting is important for
stabilization of the AFM state against doping.
We find that the structure of FSs changes significantly
with doping even in the ordered state
due to the multi-sheet FS nesting.

In the two-orbital model, we consider a square lattice of Fe ions
with $d_{zx}$ and $d_{yz}$ orbitals.~\cite{Raghu,Daghofer}
The model Hamiltonian is given by
\begin{equation}
  \begin{split}
    H=&\sum_{\mib{k},\tau,\tau^{\prime},\sigma}
    \epsilon_{\mib{k} \tau \tau^{\prime}}
    c^{\dagger}_{\mib{k} \tau \sigma}c_{\mib{k} \tau^{\prime} \sigma}
    +U \sum_{i, \tau}
    n_{i \tau \uparrow} n_{i \tau \downarrow}\\
    &+U^{\prime} \sum_{i}
    n_{i x} n_{i y}
    + J \sum_{i,\sigma,\sigma^{\prime}}
    c^{\dagger}_{i x \sigma}
    c^{\dagger}_{i y \sigma^{\prime}}
    c_{i x \sigma^{\prime}}
    c_{i y \sigma}
    \\
    &+ J^{\prime}\sum_{i,\tau \ne \tau^{\prime}}
    c^{\dagger}_{i \tau \uparrow}
    c^{\dagger}_{i \tau \downarrow}
    c_{i \tau^{\prime} \downarrow}
    c_{i \tau^{\prime} \uparrow},
  \end{split}
  \label{eq:H}
\end{equation}
where $c_{i\tau\sigma}$ is the annihilation operator of
the electron at site $i$ with orbital $\tau$
and spin $\sigma$ ($=\uparrow$ or $\downarrow$) and
$c_{\mib{k}\tau\sigma}$ is the Fourier transform of $c_{i\tau\sigma}$.
$\tau=x$ and $y$ represent $d_{zx}$ and $d_{yz}$ orbitals, respectively.
$n_{i \tau \sigma}=c^{\dagger}_{i \tau \sigma} c_{i \tau \sigma}$ and
$n_{i \tau}=\sum_{\sigma}n_{i \tau \sigma}$.
The coupling constants $U$, $U^{\prime}$, $J$, and $J^{\prime}$
denote the intraorbital Coulomb, interorbital Coulomb, exchange,
and pair-hopping interactions, respectively.
For the $t_{2g}$ orbitals, 
relations $U=U^{\prime}+J+J^{\prime}$ and $J=J^{\prime}$ hold~\cite{Tang}
and we use them.
For the kinetic energy term, we use the hopping parameters proposed
by Raghu \textit{et al.}:~\cite{Raghu}
$\epsilon_{\mib{k} xx}=-2t_1\cos k_x-2t_2\cos k_y-4t_3\cos k_x\cos k_y$,
$\epsilon_{\mib{k} yy}=-2t_2\cos k_x-2t_1\cos k_y-4t_3\cos k_x\cos k_y$,
and 
$\epsilon_{\mib{k} xy}=\epsilon_{\mib{k} yx}=-4t_4\sin k_x\sin k_y$,
where
$t_1=-t, t_2=1.3t$, $t_3=t_4=-0.85t$,
and we have set the lattice constant unity.

In this study, we consider weakly correlated cases,
e.g., $U/W \simeq 0.29$ for $U/t=3.5$, where $W=12t$ is the bandwidth.
Thus, it is reasonable to apply a Hartree-Fock approximation.
We assume that the expectation value of the number $n_{i \tau \sigma}$
is given by the following form:
\begin{equation}
  \begin{split}
    \langle n_{i \tau \sigma} \rangle
    =\{[&n+m_{s}(\delta_{\sigma \uparrow}-\delta_{\sigma \downarrow})
    +m_{o}(\delta_{\tau x}-\delta_{\tau y})\\
    &+m_{\text{so}}(\delta_{\sigma \uparrow}-\delta_{\sigma \downarrow})(\delta_{\tau x}-\delta_{\tau y})]\\
    +[&n_{\mib{q}}+m_{s \mib{q}}(\delta_{\sigma \uparrow}-\delta_{\sigma \downarrow})
    +m_{o \mib{q}}(\delta_{\tau x}-\delta_{\tau y})\\
    &+m_{\text{so} \mib{q}}(\delta_{\sigma \uparrow}-\delta_{\sigma \downarrow})(\delta_{\tau x}-\delta_{\tau y})]
  e^{i\mib{q} \cdot \mib{r}_i}\}/4,
  \end{split}
  \label{eq:HF}
\end{equation}
where $\mib{q}=(\pi,\pi)$ or $(\pi,0) \equiv \mib{Q}$,
$\mib{r}_{i}$ denotes the position of site $i$,
and $n$ is the number of electrons per site.
The order parameters are $m_s$, $m_o$, $m_{\text{so}}$, $n_{\mib{q}}$,
$m_{s\mib{q}}$, $m_{o\mib{q}}$, and $m_{\text{so}\mib{q}}$.
We determine the lowest energy state among the solutions of the Hartree-Fock approximation.
In Eq.~\eqref{eq:HF},
we consider the $z$ component for the orbital state, i.e.,
$m_o=\frac{1}{N} \sum_{i, \tau, \tau^{\prime}, \sigma} \langle c^{\dagger}_{i \tau \sigma} \hat{\tau}^{\alpha}_{\tau \tau^{\prime}} c_{i \tau^{\prime} \sigma} \rangle$
with $\alpha=z$, where $\hat{\tau}^{\alpha}$ is the Pauli matrix.
We also considered order parameters with $\alpha=x$ and $y$,
and we found that the $z$-component ordered state with $\mib{q}=\mib{Q}=(\pi,0)$
always has lower energy than the other ordered states within parameters we investigate here.

\begin{figure}
  \begin{center}
    \includegraphics[width=0.9\linewidth]{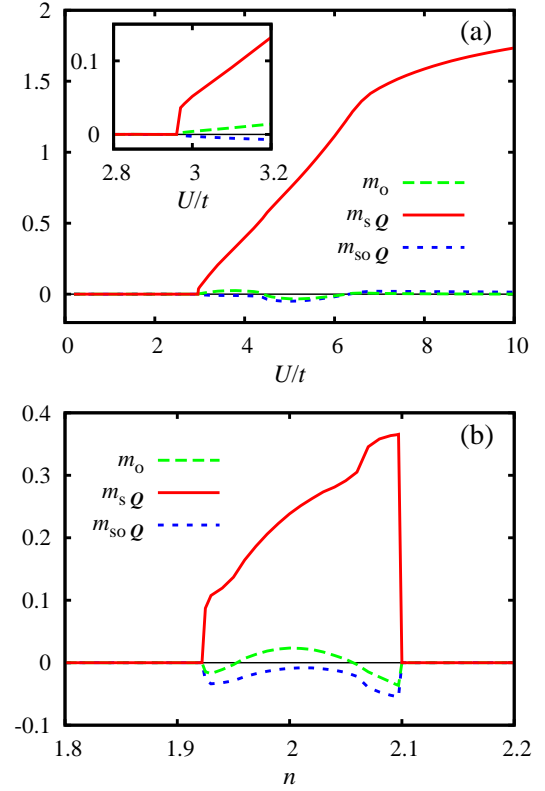}
  \end{center}
  \caption{\label{figure:orderparam}
    (Color online)
    (a) Order parameters as functions of $U$
    at $n=2$ and $J=0.1U$.
    (b) Order parameters as functions of $n$
    at $U/t=3.5$ and $J=0.1U$.
  }
\end{figure}

Figure~\ref{figure:orderparam}(a) shows $U$ dependence of the order parameters
$m_{o}$ (for FO order),
$m_{s \mib{Q}}$ (for AFM order), and
$m_{\text{so} \mib{Q}}$ (for antiferro-spin-orbital order)
at $n=2$ and $J=0.1U$.
We find that the other order parameters are zero.
As shown in the inset, $m_{s \mib{Q}}$ jumps to a finite value at $U/t \simeq 2.97$,
and $m_{o}$ and $m_{\text{so} \mib{Q}}$ also have jumps to finite values
at the same point while they are small and not visible on the scale of Fig.~\ref{figure:orderparam}(a).
Thus, the transition to the AFM state is of first order.

Figure~\ref{figure:orderparam}(b) shows doping dependence of
the order parameters at $U/t=3.5$ and $J=0.1U$.
We have chosen this value of $U$
so as to the AFM state is destabilized
by the doping of $\sim 0.1$
as in experimental observations.~\cite{Kamihara,Rotter_dope2,Chu,Luetkens}
The AFM moment changes continuously with doping at first,
but suddenly disappears at $n \simeq 1.92$ and 2.1.
%
We obtained small values of $m_{s \mib{Q}}$, which are much smaller than
the saturation value $n$ (for $n \le 2$) or $4-n$ (for $n > 2$),
as in experimental observations ($m_{s \mib{Q}}=1$ is corresponding to 1~$\mu_{\text{B}}$ of an AFM moment):
0.25~$\mu_B$,
0.35~$\mu_B$, or
0.36~$\mu_B$ in LaFeAsO;~\cite{Luetkens,Cruz,Kitao,Klauss}
0.94~$\mu_B$ or
1.01~$\mu_B$ in SrFe$_2$As$_2$;~\cite{Zhao,Kaneko}
0.4~$\mu_B$ or
0.87~$\mu_B$ in BaFe$_2$As$_2$.~\cite{Huang,Rotter_undope}
The estimated values of the ordered moments
depend on the experimental probes
even for the same material,
and the reason of this discrepancy is not clear at present.
In the electron doped case,
the AFM moment increases with doping,
and the transition temperature is expected to become higher
than the undoped case.
It is in contradiction to experimental observations,
and to resolve this discrepancy, we have to extend the model,
e.g., by using the five-orbital basis.

Note that in the AFM state with $\mib{Q}=(\pi,0)$,
$x$ and $y$ directions are not equivalent,
and the occupancies of $d_{zx}$ and $d_{yz}$ become different.
Thus, the AFM state in the multiorbital system
inevitably accompanies FO order, i.e., finite $m_o$.
Through an electron-lattice interaction,
the FO order results in a lattice distortion
from a tetragonal to orthorhombic structure.
This is consistent with experimental observations that
the AFM phase is always orthorhombic.
The obtained small values of the order parameter $m_{o}$
for the FO order may be responsible for
the weakness of anomaly in lattice distortion, e.g.,
small volume change in SrFe$_2$As$_2$ at the transition.~\cite{Tegel}
Note that another scenario is proposed for the lattice distortion,
in which the lattice distortion relaxes magnetic frustration
and is necessary for occurrence of the AFM order.~\cite{Yildirim}
On the contrary, in our theory,
the lattice distortion is a secondary effect due to the FO order accompanied by the AFM order.
Since the AFM state in Fe pnictides is metallic,
we believe that our picture is more suitable for Fe pnictides.
In the coexistent state of antiferromagnetism and FO order,
$m_{\text{so} \mib{Q}}$ also becomes finite as shown in Fig~\ref{figure:orderparam}.
Note that we obtain similar results for $J=0$
at least in a small-$U$ region and the choice of the value of $J$
does not change the present results qualitatively.
In this model, at zero temperature,
the AFM order and the FO order disappear at the same doping
and it is consistent with the experimental observations.
However, some extensions, e.g., inclusion of electron-lattice interaction,
may be necessary to obtain different transition temperatures
for the AFM order and for the FO order as in LaFeAsO.

\begin{figure}
  \begin{center}
    \includegraphics[width=\linewidth]{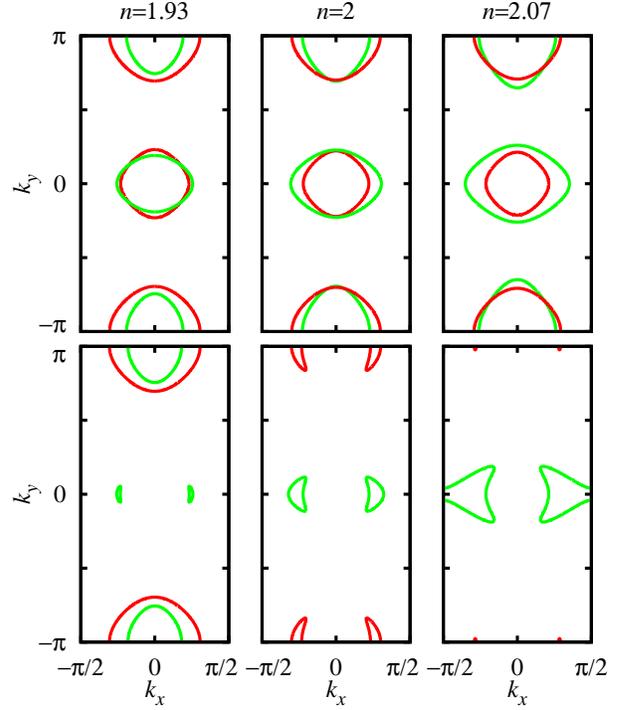}
  \end{center}
  \caption{\label{figure:FS}
    (Color online)
    FSs in the folded BZ
    with respect to $\mib{Q}=(\pi,0)$
    for $n=1.93$, 2, and 2.07.
    The upper panels show the FSs
    in the normal state.
    The lower panels show those in the AFM state
    with FO order at $U/t=3.5$ and $J=0.1U$.
  }
\end{figure}

To obtain further insights into the ordered states,
we show FSs in Fig.~\ref{figure:FS}
in the normal and ordered states
at half-filling ($n=2$) and at around phase boundaries.
It is evident from the figure that the $x$ and $y$ directions are not equivalent
in the ordered states.
In the hole doped case, $n=1.93$,
nesting between FSs centered at $\mib{k}=(0,0)$ and $(\pi,0)$
in the unfolded BZ is strong.
Then,
these FSs are reconstructed into small pockets around $(\sim\pm\pi/4,0)$,
while the other FSs centered at $(0,\pi)$ and $(\pi,\pi)$ are almost unchanged.
On the other hand, in the electron doped case, $n=2.07$,
nesting is strong between FSs centered at $\mib{k}=(0,\pi)$ and $(\pi,\pi)$.
Around zero doping, both types of nesting can contribute to stabilize
the AFM state.
As a result,
the doping effect on the AFM moment is not so significant
in the lightly doped cases,
while the structure of the FSs changes very much.
Thus, the multi-sheet FS nesting is important for the stabilization
of the AFM state in this system.
When we dope carriers further, the nesting becomes weak
and the AFM state is destabilized suddenly.

Note that such a mechanism to stabilize the AFM state
against doping is applicable
as long as the sizes of the hole surfaces around $(0,0)$ and $(\pi,\pi)$
are different and the undoped system is a compensated
or nearly compensated metal.
The smaller hole surface mainly contributes for the realization of
antiferromagnetism for the hole doped case,
and the larger hole surface mainly contributes for the electron doped case.
Thus, this mechanism works irrespective of precise choice of the model parameters.
We also note that in a more realistic model, i.e., five-orbital model,~\cite{Kuroki}
both the two hole surfaces locate around $(0,0)$ even in the unfolded BZ.
However, the nesting vector $\mib{Q}=(\pi,0)$ is the same as in the two-orbital
model, and the present mechanism to stabilize the AFM state is applicable
provided the two hole surfaces have different sizes.

At $n=2$ only small pockets of FSs remain in the ordered states.
The area of one pocket at $n=2$ is 0.86\%
of the folded BZ for $\mib{Q}=(\pi,0)$
[and of the normal state BZ folded due to
the actual lattice structures of Fe pnictides
(two Fe ions per unit cell)].
There are two electron pockets and two hole pockets,
but the areas of them are the same,
since the model is a compensated metal at $n=2$
and the two electron (hole) pockets occupy the same amount of area due to symmetry.
Experimentally observed volumes of FSs
in the AFM state are small:
0.26\%-1.38\% in SrFe$_2$As$_2$~\cite{Sebastian} and
0.3\%-1.7\%   in BaFe$_2$As$_2$~\cite{Analytis} of the normal state folded BZ.
These values are comparable with our theoretical ones.
In the normal state at $n=2$,
the hole pocket around $(0,0)$,
hole pocket around $(\pi,\pi)$, and
electron pocket around $(\pi,0)$
occupy 7.13\%, 13.24\%, and 10.18\% of the folded BZ, respectively.
These values are also comparable with experimental ones,
2.8\%-9\% of BZ in LaFePO in the normal state.~\cite{Coldea}

\begin{figure}
  \begin{center}
    \includegraphics[width=\linewidth]{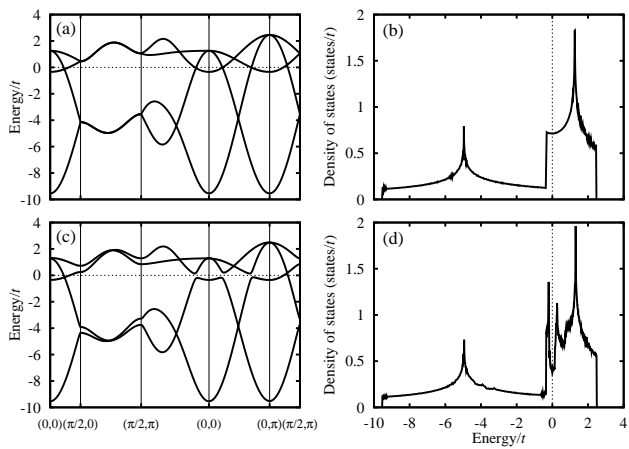}
  \end{center}
  \caption{\label{figure:DOS}
    Band structure in the folded BZ for $\mib{Q}=(\pi,0)$
    and density of states at $n=2$.
    (a) Band structure and (b) density of states in the normal state.
    (c) Band structure and (d) density of states
    in the AFM state with FO order
    at $U/t=3.5$ and $J=0.1U$.
    The Fermi energy is set to be zero in these figures.
  }
\end{figure}
Figure~\ref{figure:DOS} shows the band structure and density of states
in the normal and ordered states.
In the ordered state, band gaps open at some points at the Fermi level,
while not at $(\sim \pi/4,0)$ and $(\sim \pi/4,\pi)$.
At $k_y=0$ and $\pi$,
the off-diagonal element $\epsilon_{\mib{k} xy}$ in the kinetic energy term is zero,
and $d_{zx}$ and  $d_{yz}$ orbitals do not mix.
The mean field in the ordered state mixes electrons
with $\mib{k}$ and $\mib{k}+\mib{Q}$ in the same orbitals,
and $d_{zx}$- and $d_{yz}$-orbital states are not mixed
even in the ordered state at $k_y=0$ and $\pi$.
The two bands crossing at around the Fermi level
are different orbitals
at both $(\sim \pi/4,0)$ and $(\sim \pi/4,\pi)$,
and a gap cannot open there.
As a result, the FSs do not disappear
on the lines $k_y=0$ and $\pi$
even in the AFM state as shown in Fig.~\ref{figure:FS}.
Note that in Fig.~\ref{figure:FS} the Fermi pockets at $n=2.07$ on $k_y=\pi$ are very small
in the ordered state but have finite volumes.
Thus, the system remains metallic in the AFM state
as in experimental observations.
The density of states in the ordered state has a gap-like structure
around the Fermi level, but remains finite at the Fermi level.
For larger $U$ cases, the band structure changes very much,
and the system can become insulating.~\cite{Yu}
By increasing $U$, the band crossing points on $(0,0)$-$(\pi/2,0)$
and $(0,\pi)$-$(\pi/2,\pi)$ lines move to $(\pi/2,0)$ and $(\pi/2,\pi)$,
respectively, and finally the band crossing disappears.
Then, the system can become insulating, e.g., at $U/t \gtrsim 6.7$
for $n=2$ and $J=0.1U$.
Thus, $U$ should not be very large in Fe pnictides.

In conclusion, we have shown that characteristic features
of the AFM state in Fe pnictides
can be naturally understood within the two-orbital model.
The stability of AFM phase is due to the
multi-sheet FS nesting.
The tetragonal to orthorhombic lattice distortion is
a secondary effect due to the FO order
but not a driving mechanism of antiferromagnetism.
FSs remain in the ordered state
due to the multiorbital character of the crossing bands.
Our theory indicates that the FSs change significantly upon doping.
In the doped AFM states around phase boundaries,
some Fermi pockets become very small, while the other Fermi pockets
have large volumes as in the normal state.
Experimental observations of these FSs are highly desired,
since we can know what kind of nesting is strong around the phase boundaries
from the reconstructed FSs.
Such a knowledge is important to unveil fluctuations
which mediate the superconducting pairing.

\section*{Acknowledgments}
KK thanks K. Kaneko and A. Moreo for useful comments.
KK is supported by Japan Society for the Promotion of Science
through a Postdoctoral Fellowship for Research Abroad.


\end{document}